# The New Pro Se:<br>Generative AI and the Surge in Federal Civil Self-Representation

**Or Cohen-Sasson**
Miami Law & AI Lab, University of Miami School of Law
orcs@law.miami.edu

**Abstract**

Since public access to generative AI tools became widespread, federal civil litigation has seen a marked increase in pro se (self-represented) plaintiffs. This paper analyzes that shift using ~2.8 million filings, asking whether the post-GenAI period is associated not only with more pro se filings, but also with detectable changes in complaint text, litigation outcomes, and the composition of pro se litigants.

Using civil filing data from FY2008–2025, we find that the federal civil pro se plaintiff rate rose from 11.33% pre-GenAI to 16.94% post-GenAI, a 5.61 percentage-point increase that persists after trend and covariate-adjusted robustness checks. We then focus on Civil Rights and Other Statutory cases, where the increase is especially pronounced, and link case metadata to pro se complaints. Drawing on stylometric AI detection indicators, we develop an interpretable measure of AI-consistent drafting. Against a threshold calibrated to the pre-GenAI baseline, the net AI-flagged share is 13.9% of post-GenAI non-form complaints.

Analysis of the AI-flagged complaints shows that they are more citation-dense, disproportionately associated with first-time rather than repeat filers, and geographically unevenly distributed. This composition pattern suggests that AI-consistent drafting is not merely a repeat-filer phenomenon; it also includes a modest, suggestive increase in name-inferred female plaintiffs. We find no evidence of improved win rates; in fact, AI-flagged complaints are more likely to be dismissed and to terminate at earlier procedural phases. These findings raise new questions about access to justice and court screening burdens, and sharpen the distinction between legal formality and legal efficacy.

## 1. Introduction

Since late 2022, generative AI (GenAI) has diffused rapidly, reshaping professional and everyday activity (Bick, Blandin, and Deming 2024). In law, much of the debate over AI has focused on professional uses: how lawyers use AI (McGinnis and Pearce 2014; Susskind and Susskind 2022), how courts should regulate AI-generated filings (New York State Bar Association 2024; Charlotin 2026), or whether legal services will be automated (Hadfield 2017; Barton 2015). But GenAI may also matter before any lawyer or judge becomes involved. It may change the threshold act by which a person chooses to turn a grievance into a lawsuit (Felstiner, Abel, and Sarat 1981).

That threshold is especially important when people file suit without a lawyer ("pro se" or self-represented plaintiffs). Pro se plaintiffs must produce legal documents that are difficult to draft without professional expertise (Hammond 2022). If GenAI lowers that drafting cost, time, and expertise, the federal civil docket is one of the first places where the institutional consequences of broad GenAI access should become visible.

Anecdotal reports of pro se litigants using GenAI have circulated since 2023 (Volokh 2023). But there is no large-scale empirical evidence on whether the post-2022 diffusion of GenAI is associated with measurable changes in who files federal civil cases, what their complaints look like, and what happens to those cases after they are filed. This paper provides that evidence.

We use a two-stage empirical design covering FY2008–FY2025. The first stage analyzes approximately 2.8 million federal civil filings from the Federal Judicial Center Integrated Database (FJC). The second examines 12,842 pro se complaints from CourtListener, linked back to FJC metadata and focused on Civil Rights and Other Statutory, the two categories where the macro signal is strongest.

Three findings stand out. First, the plaintiff pro se rate rose from 11.33% pre-GenAI to 16.94% post-GenAI—a 5.61 percentage-point increase, or roughly a 50% relative change—and interrupted time-series evidence suggests an accelerating post-AI trend rather than a one-time level shift, robust to district, case-category, and COVID-era controls. Second, in the linked complaint-level sample, AI-flagged complaints appear more formally polished than unflagged complaints—they are markedly more citation-dense—yet they perform worse on some case trajectories: higher dismissal rates (61.1% vs. 53.6%), more terminations at early procedural stages, and no win-rate advantage. Third, the post-AI shift is unevenly distributed: AI-flagged complaints concentrate in specific federal circuits and are substantially more likely than unflagged complaints to be associated with observed first-time filers. That pattern weakens a pure serial-litigator account of the AI-flagged subset. We also observe a modest, borderline-significant increase in name-inferred female plaintiffs among AI-flagged non-form complaints. At the same time, the pro se surge does not map cleanly onto the financially constrained populations most often centered in access-to-justice discussions.

The paper contributes to the study of AI and society by examining public GenAI adoption in the concrete institutional setting of federal civil courts. The findings complicate simple accounts of technology-enhanced access-to-justice. GenAI may help more people produce complaints and enter court, but the resulting access appears uneven and does not necessarily translate into direct legal success. This has implications not only for access-to-justice debates, but also for court administration. If litigants can adopt public GenAI tools faster than courts can adapt to increased filing and screening burdens, AI may reshape legal institutions from the bottom up, even without formal court adoption.

# 2. Background and Research Questions

## 2.1 Background and Related Work

The dominant scholarly conversation about AI and law has been overwhelmingly lawyer-side and court-side. It examines how AI reshapes legal research (Magesh et al. 2025), document review (Grossman and Cormack 2011), and litigation prediction (Aletras et al. 2016); whether it may displace or restructure professional legal work (Remus and Levy 2017); and how legal technology more broadly is reshaping civil procedure and the adversarial process (Engstrom and Gelbach 2021). The litigant-side question this paper takes up, that is, whether public access to GenAI tools is associated with changes in what laypeople do at the threshold of court, has received less systematic attention.

That question matters because self-representation is a persistent and structural feature of U.S. civil litigation, not a marginal one (Swank 2005). Pro se litigants have consistently accounted for roughly one in nine non-prisoner federal civil cases (Gough and Taylor Poppe 2020) and they face well-documented disadvantages in navigating substantive law and procedure (Carpenter et al. 2018). Pro se litigation sits at the heart of long-running access-to-justice debates, which document a persistent gap between civil legal needs and available legal services (Sandefur 2016; Engstrom and Engstrom 2024) and treat self-representation as both a symptom of that gap and a site for institutional response (Rhode 2004). A long-standing thread in this literature distinguishes formal entry into the legal system from substantive access to legal remedies (Sandefur 2019). Technology occupies a contested place in this debate. Optimists argue that legal technology, and now generative AI, can lower the cost of drafting and procedural navigation and thereby democratize entry into courts (Barton and Bibas 2017; Susskind and Susskind 2022; McGinnis and Pearce 2014). Skeptics warn that diffusion is uneven, that AI may produce a two-tiered system of legal help, and that surface-level legal assistance does not reliably translate into substantive legal gains (Simshaw 2022; Greiner and Pattanayak 2012; Greiner, Pattanayak, and Hennessy 2013).

Running parallel to this access debate is a separate but related question about lawyer displacement: whether AI substitutes for human legal labor. Pre-LLM evidence suggested that displacement would be task-specific and partial rather than wholesale (Remus and Levy 2017). A pro se-side analysis provides one type of partial evidence on this question, by showing whether represented filings actually decline when self-represented filings rise.

Practitioner reports since 2023 describe AI-assisted pro se filing as an emerging and consequential phenomenon (Strom 2026; Lex Machina 2026; Poggio 2024; Charlotin 2026). However, systematic empirical evidence at the federal-docket scale has not been available.

## 2.2 Study Scope and Research Questions

Against this background, this work asks whether the public diffusion of (GenAI), and specifically large-language models (LLMs), is associated with measurable changes in federal civil self-representation. The study examines a concrete institutional setting in which public-facing GenAI tools might plausibly matter: the initiation of federal civil litigation by self-represented plaintiffs. Federal complaints are a useful site for this inquiry because they are both legally consequential and practically difficult for nonlawyers to draft (Reinert 2015; Hatamyar Moore 2010). If GenAI reduces the cost, time, and expertise required for pro se litigation, one place this should appear is at the threshold of litigation, that is, filing a civil complaint: the volume of filings, who files, what they look like, and what happens to them after filing.

The empirical analysis proceeds in two stages. First, the study asks whether the rate of pro se federal civil complaints changed after public access to GenAI tools became widespread in November 2022. This is the macro filing question, which establishes the baseline: is there a post-AI pro se filing shift, and is it concentrated in particular case categories?

Second, the study asks whether the post-AI filing shift, if it exists, is accompanied by detectable changes in complaint drafting, plaintiff composition, case outcomes, and other litigation characteristics. This is the complaint-level question. Because the study does not observe individual AI use directly (i.e., no litigant survey was conducted), it develops a measure of AI-consistent drafting signatures in the complaint-text sample. The analysis then focuses on two focal case categories—Civil Rights and Other Statutory cases—because they (1) show especially pronounced post-AI increases in pro se filing rates, and (2) are large federal civil categories, together representing approximately 36.7% of all-civil filings. Within those categories, the study distinguishes between *non-form complaints* and *form complaints*. The latter class serves as a quasi-control group within the post-AI period; Section 3 explains this design rationale.

This design yields two research questions:

**Research Question 1 (RQ1): Macro filing change.** How did federal civil plaintiff pro se rate change after public access to GenAI tools became widespread?

**Research Question 2 (RQ2): AI-consistent complaint drafting and associated litigation patterns.** Among pro se complaints in the focal case categories, what share of post-GenAI complaints exhibit strong AI-consistent drafting signatures? How, if at all, do AI-flagged complaints differ from the unflagged comparison group in drafting, filer composition, and litigation success?

Together, these questions separate the paper's two claims. RQ1 concerns magnitude: whether federal civil self-representation grew unusually fast after public access to GenAI tools became widespread, and whether any such growth is concentrated in particular case categories. RQ2 concerns composition: whether a meaningful share of post-AI pro se complaints carry textual signatures consistent with GenAI drafting, and whether those AI-consistent complaints exhibit distinct patterns and trajectories.

Linking RQ1 and RQ2 matters because a rise in pro se filing is, on its own, an ambiguous signal: it could reflect a narrow set of serial litigants filing more, or genuine broadening of participation that nonetheless yields only *access to drafting* rather than *access to remedies*. The paper's contribution is an integrated strategy linking macro trends to complaint text, composition, and outcomes.

# 3. Data and Empirical Design

This study uses three nested data layers. The first is a filing-level dataset used to measure changes in all-civil filings. The second is a complaint-text sample used to measure AI-consistent drafting signatures. The third is a linked complaint sample that connects complaint text to civil filings metadata, enabling analysis of case-level characteristics. The design is descriptive and associational: it identifies temporal shifts and complaint-level patterns, rather than causal effects.

## 3.1 Filing Data, Period, and Analytic Universes

The filing analysis uses civil case records from the FJC. The analysis loads fiscal-year files from FY2008 through FY2025. We use November 30, 2022, the public release of ChatGPT (OpenAI 2022), as the GenAI cutoff. Cases are assigned to periods by filing date.

For RQ1, the paper studies all-civil filings, that is, all original federal civil filings during FY2008-FY2025, excluding prisoner petitions and filings with unknown case types. The all-civil filing universe contains approximately 2.8 million filings and is used to establish the overall post-AI change in federal civil self-representation. Cases are treated as plaintiff pro se when the plaintiff is self-represented, regardless of the representation status of the other party. Cases coded as represented are used as a comparison group in several robustness and controlled models. This broad universe establishes the macro trend.

Based on RQ1 findings, for RQ2, the paper narrows to two focal case categories: Civil Rights and Other Statutory. These categories are appropriate for deeper complaint-level analysis because, as Section 4 shows, both exhibit especially pronounced post-AI increases in pro se rates. The focal categories are not a niche but rather two of the most substantial categories, accounting for ~37% of the all-civil filing universe. The focal categories therefore provide an empirically powered and substantively important setting for studying the complaint-level question: if GenAI is connected to the post-2022 pro se filing shift, these are the categories where that connection should be most visible.

The main filing analyses combine simple pre/post comparisons and quarterly models. Quarterly trajectories, interrupted time-series models, and covariate-adjusted logistic regressions test whether filing shifts are large, temporally concentrated, and robust to pre-existing trends, COVID-era disruption, case-category composition, and district-level differences.

## 3.2 Complaint-Text Sample and Form/Non-Form

The complaint-level analysis uses pro se complaints in the focal case categories. This sample is drawn from complaint documents filed during FY2018-FY2025 available through CourtListener. After text preparation and length filters, the complaint-text sample contains 12,842 pro se complaints. The sample is narrower than the all-civil filing universe because it includes only cases for which complaint text is available.

For complaint-level analyses, the study distinguishes between non-form and form complaints. Form complaints are submitted on standardized court templates that streamline pro se drafting (Greiner, Pattanayak, and Hennessy 2013); non-form complaints are drafted freely. Direct, whole-document AI-assisted drafting is both more likely to occur and more detectable in non-form complaints: free-prose drafting space invites copy-paste use, and the non-template nature of non-form complaints exposes AI-oriented drafting signatures that template language in form complaints can dilute or obscure. The expectation is not that form complaints will show no post-AI movement; AI use there may simply be more partial, consultative, or obscured.

We therefore treat form complaints as a quasi-control group rather than a negative control: post-AI form complaints may exhibit AI assistance, but the detectable signal is expected to be weaker.

### 3.3 Measuring AI-Consistent Drafting

The paper does not observe individual GenAI use; instead, it constructs an interpretable, pre-AI-calibrated proxy for AI-consistent drafting from three textual indicators drawn from existing literature. Section 5 details the indicators, threshold, and credibility checks; the relevant signal is the net post-AI increase in AI-flagged complaints relative to the pre-AI baseline.

### 3.4 Linking Complaints to Case-Level Measures

The final stage links the complaint-level dataset to the FJC dataset. The linked complaint sample consists of complaints from the complaint-text sample that can be matched to FJC records using case identifiers. The linked data support connecting complaints to case-level measures and comparing them across drafting features, filer composition, case outcomes, and more; Section 6 reports the specific measures analyzed. The main comparison is between the AI-flagged subset (i.e., post-AI non-form complaints that exceed the AI-score threshold) and the unflagged comparison group (i.e., post-AI non-form complaints below the threshold). Additional analyses use form complaints, pooled form-and-non-form samples, and pre-AI high-score complaints to assess robustness and baseline differences.

First-time/repeat filer status is based on primary-plaintiff names from the focal sample. A case is coded as observed first-time when the normalized primary plaintiff name first appears in that sample; this is not a claim that the person had never filed in federal court, in another case category, in an unlinked case, or in another court system. Name-inferred gender is estimated from the primary-plaintiff name. Gender is not self-reported; the main analysis uses NamSor (Sebo, Shamsi, and Wang 2026) with prediction probability threshold >= 0.80 for binary classification.

## 4. More Plaintiffs Without Lawyers

RQ1 asks whether federal civil self-representation changed after public access to GenAI tools became widespread. The answer is yes. Across all-civil filings, the post-AI period is associated with a sharp increase in the share of cases filed by self-represented plaintiffs. The increase appears not only in the full pre/post comparison, but also in quarterly trajectories, trend models, and case-category-specific analyses. The surge is not evenly distributed across the federal docket: the strongest increases appear in particular categories, including the Civil Rights and Other Statutory categories that become the focus of the complaint-level analysis in Sections 5-6.

### 4.1 A Sharp Increase in All-Civil Pro Se Filings

The all-civil filing analysis shows a large post-AI increase in plaintiff self-representation. The all-civil plaintiff pro se rate rose from 11.33% before the GenAI cutoff to 16.94% after it, a +5.61 percentage-point increase, or a 49.5% relative increase. The case-level difference is statistically significant ($\chi^2$=10,478.21, p<0.001).

The same pattern holds when summarized by quarter. A comparison of the last four pre-AI quarters to the last four post-AI quarters yields an even larger gap (13.06% vs. 19.59%, +6.53 pp), and by FY2025-Q4 pro se filings reached 22.30% of all-civil filings.

The filing-count evidence shows that this is not only a rate phenomenon. Average quarterly all-civil counts rose modestly from 36,697 to 37,765 (+2.9%). Over the same period, average quarterly plaintiff pro se filings rose from 4,152 to 6,423 (+54.7%), while represented filings fell slightly, from 31,654 to 30,637 (-3.2%). Thus, the post-AI period is characterized by a substantial increase in filings by self-represented plaintiffs, not merely by a uniform expansion of the federal civil docket.

The count data do not provide clear evidence of lawyer displacement: in Civil Rights, both pro se and represented filings rose; in Other Statutory, the represented-case decline pre-dates the GenAI cutoff.

### 4.2 Acceleration, not a One-time Level Shift

The all-civil trend evidence suggests that the post-AI pattern is better described as post-AI acceleration than as a single jump. Before the GenAI cutoff, the plaintiff pro se filing-rate trend was nearly flat: +0.078 percentage points per year, with only borderline statistical evidence (p=0.051). After the cutoff, the trend steepened sharply to +2.996 percentage points per year (p<0.001).

Interrupted time-series models support the same interpretation (Lopez Bernal, Cummins, and Gasparrini 2017). In the all-civil model, the immediate level shift at the GenAI cutoff is positive but not statistically significant (+1.090 percentage points, p = 0.171). The post-AI slope change, however, is large and statistically significant: +0.729 percentage points per quarter (p<0.001). A COVID-controlled specification produces the same substantive result, with a post-AI slope change of +0.727 percentage points per quarter (p<0.001). At the final observed period, the ITS-modeled cumulative effect is approximately eight percentage points above the pre-AI trend projection.

Covariate-adjusted models support the same conclusion. A logistic regression of plaintiff pro se status on the post-AI period, controlling for district and NOS group, estimates a post-AI odds ratio of 1.364 (95% CI [1.351, 1.377],

p<0.001). Adding a COVID FY2020 control leaves the result essentially unchanged (OR = 1.352, 95% CI [1.339, 1.366], $p < 0.001$). Notably, these models do not identify GenAI as the causal mechanism for the pro se surge. They do, however, help rule out competing explanations (e.g., district composition or broad case-category mix).

Taken together, the descriptive and modeled evidence supports a cautious but clear conclusion: after public GenAI became widely available, federal civil plaintiff pro se filing increased sharply, and the increase appears as an accelerating post-AI trend rather than a continuation of historical patterns. The shape is consistent with a gradual technology diffusion process (Rogers 2003): if public GenAI tools affected filing behavior, their influence would likely grow over time as awareness spread.

## 4.3 Case-category Heterogeneity

The all-civil result masks substantial variation across types of civil cases. Some case categories show large post-AI increases in plaintiff pro se filing rates; others show smaller increases, near-flat patterns, or declines. This heterogeneity matters for two reasons. First, it guards against an overly broad story in which all civil litigation is assumed to respond to GenAI in the same way. Second, it identifies where complaint-level analysis is most likely to be informative and motivates the narrower design for Sections 5 and 6.

The largest increases are concentrated in Other Statutory and Civil Rights. Other Statutory cases show the largest exact-date increase, rising from 11.23% pre-AI to 19.51% post-AI (+8.28 percentage points, p<0.001). Civil Rights cases also increase sharply, from 28.16% to 35.60% (+7.44 percentage points, p<0.001). Other increases appear in Real Property (+7.23), Torts (+5.23), Tax (+4.67), and Contract (+3.28). Near-flat pattern is visible in Forfeiture/Penalty (+1.08), Labor (+0.61), and Intellectual Property (+0.14), while Social Security (-2.31) and Immigration (-3.73) decline. All are significant at p<0.001, except Intellectual Property (p=0.064).

This heterogeneity is important because it suggests that any GenAI-related filing shift, if present, is unlikely to operate uniformly across the federal docket. Several mechanisms could produce such variation. One is claim stakes: categories differ in economic value and practical consequences, which may affect whether plaintiffs seek counsel or file on their own. Social Security cases, for example, often involve monetary benefits (Hoynes, Maestas, and Strand 2022), while Immigration cases may involve high-stakes practical consequences (Eagly and Shafer 2015) that make self-representation dynamics different from lower-stakes civil claims. Another mechanism is legal specialization: highly technical domains, such as Intellectual Property, may remain less susceptible to self-represented filing even if general-purpose drafting tools become widely available (Gaudry 2012). These explanations are suggestive rather than exhaustive; lawyer availability, legal costs, and other category-specific dynamics may also shape filing patterns.

## 4.4 The Focal Case Categories

The same heterogeneity helps discipline the paper's next stage. Since the filing surge is uneven, the complaint-level analysis should focus on categories where the macro signal is strong enough to support testing for AI-consistent drafting and downstream litigation patterns. Averaging across categories where the relevant signal may be weak, absent, or opposite may dilute the phenomenon the complaint-level analysis is designed to study.

Two considerations make Civil Rights and Other Statutory the appropriate focus. Both show pronounced post-AI pro se increases (Civil Rights: from 28.16% to 35.60%; Other Statutory: from 11.23% to 19.51%), and together they account for approximately 36.7% of all-civil filings—major parts of the federal civil docket rather than niche categories.

Solo-category analyses reinforce this choice. In Civil Rights, the post-AI pattern is especially striking because it reverses a declining annual pre-AI trend (-0.413 percentage points) to substantial positive trend (+4.745 percentage points). Other Statutory has the largest exact-date increase among all categories, and the interrupted time-series model shows both a significant level shift and significant slope change. These results make Civil Rights and Other Statutory an empirically powered and substantively meaningful setting for the complaint-level analyses that follow.

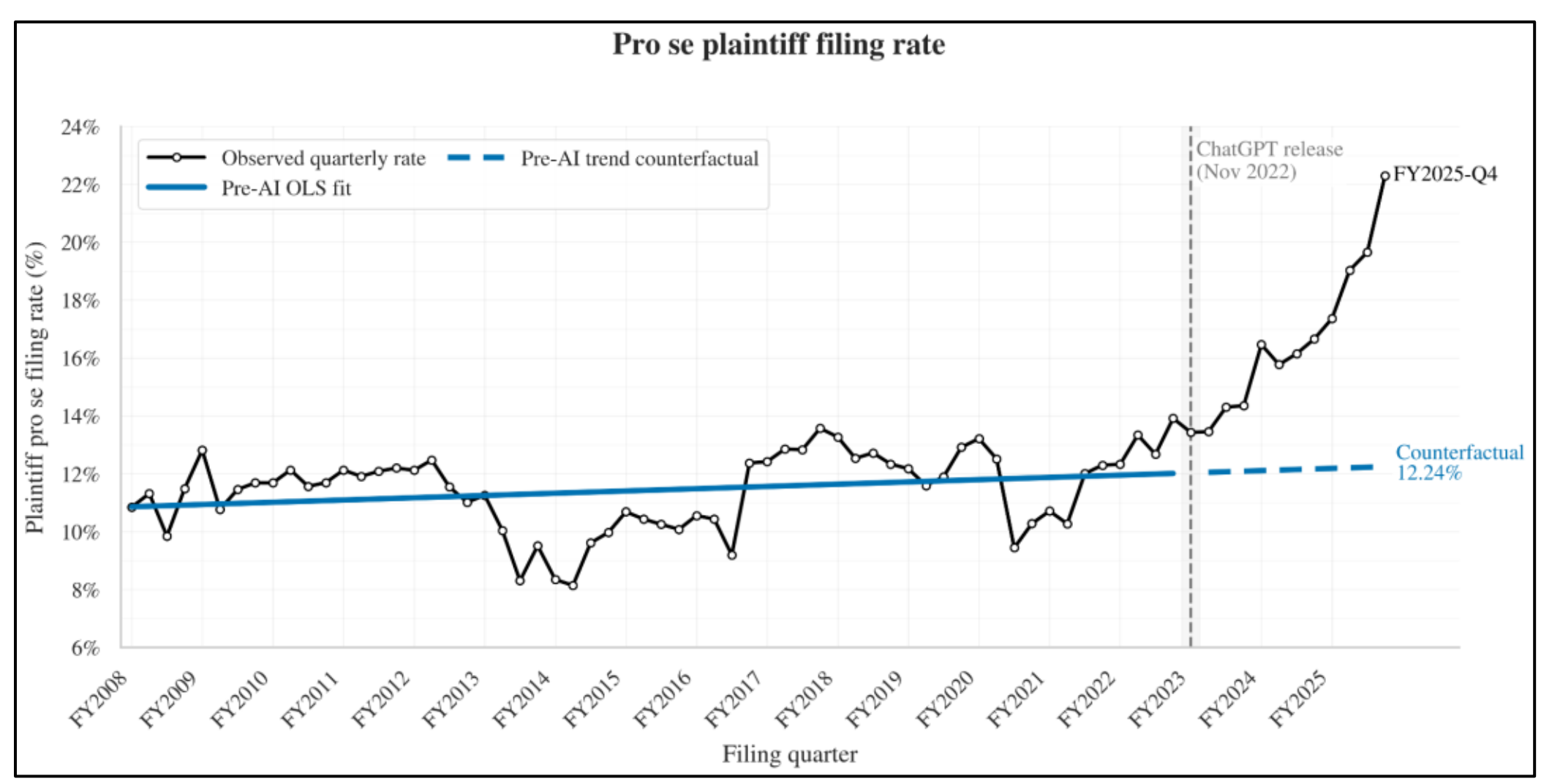

*Figure 1. Quarterly pro se plaintiff filing rate in all-civil federal filings, FY2008-Q1 to FY2025-Q4 (black). Solid blue: OLS fit to pre-GenAI quarters (through FY2022-Q4); dashed blue: counterfactual extrapolation. FY2023-Q1 (containing the November 30, 2022 ChatGPT release) is the transition quarter. The exact-date rate rose from 11.33% to 16.94% (+5.61 pp, p<0.001); FY2025-Q4 reached 22.30%.*

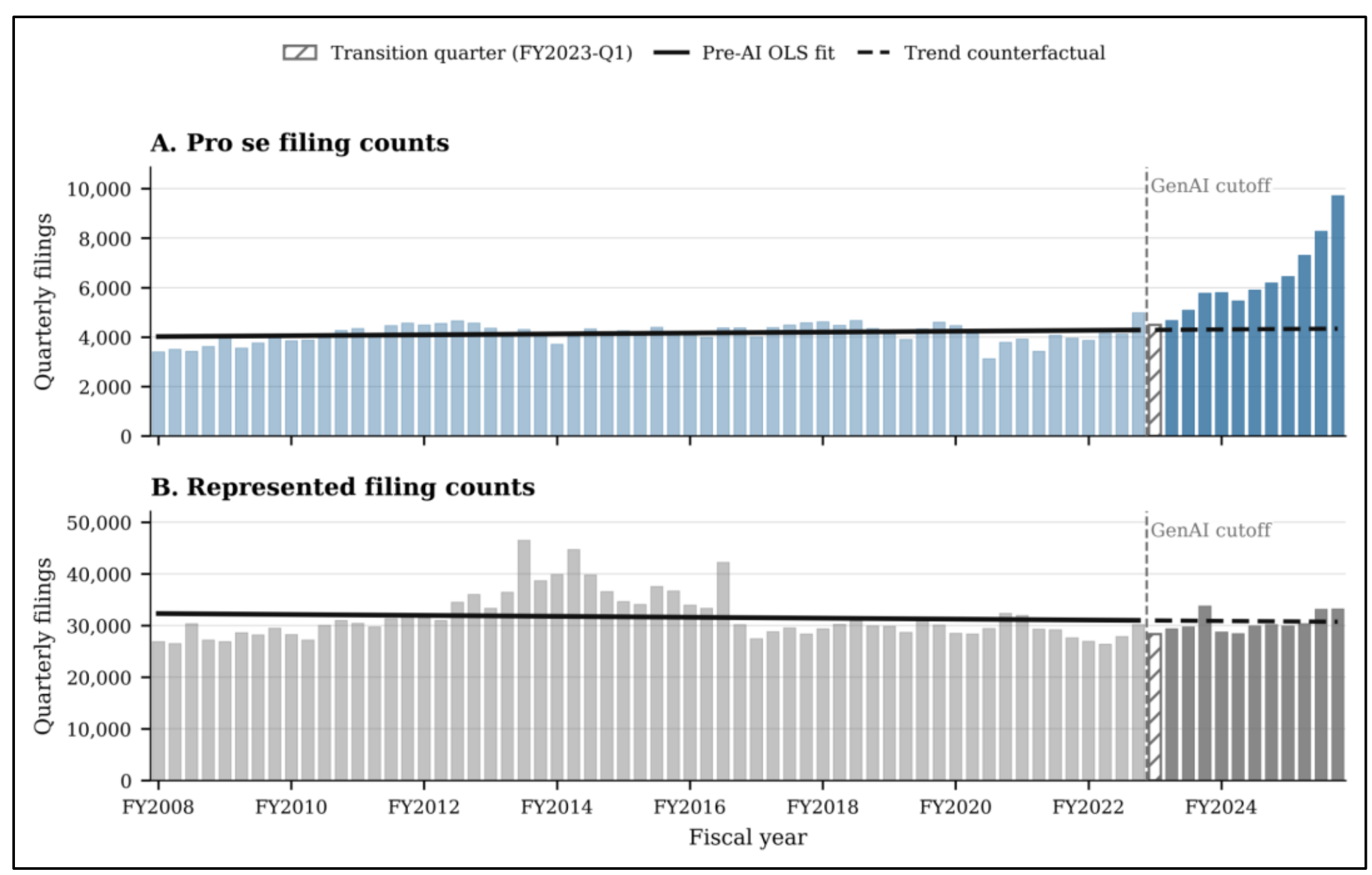

*Figure 2. Quarterly pro se plaintiff (Panel A, blue) and represented (Panel B, gray) federal civil filing counts, FY2008-Q1 to FY2025-Q4. Hatched bar: FY2023-Q1 transition quarter (Nov. 30, 2022 ChatGPT release). In each panel, solid line: OLS fit to pre-AI quarters; dashed line: counterfactual extrapolation. After the cutoff, average quarterly pro se filings rise from 4,152 to 6,423 (+54.7%), well above the counterfactual; represented filings move from 31,654 to 30,637 (-3.2%), showing no comparably sharp departure.*

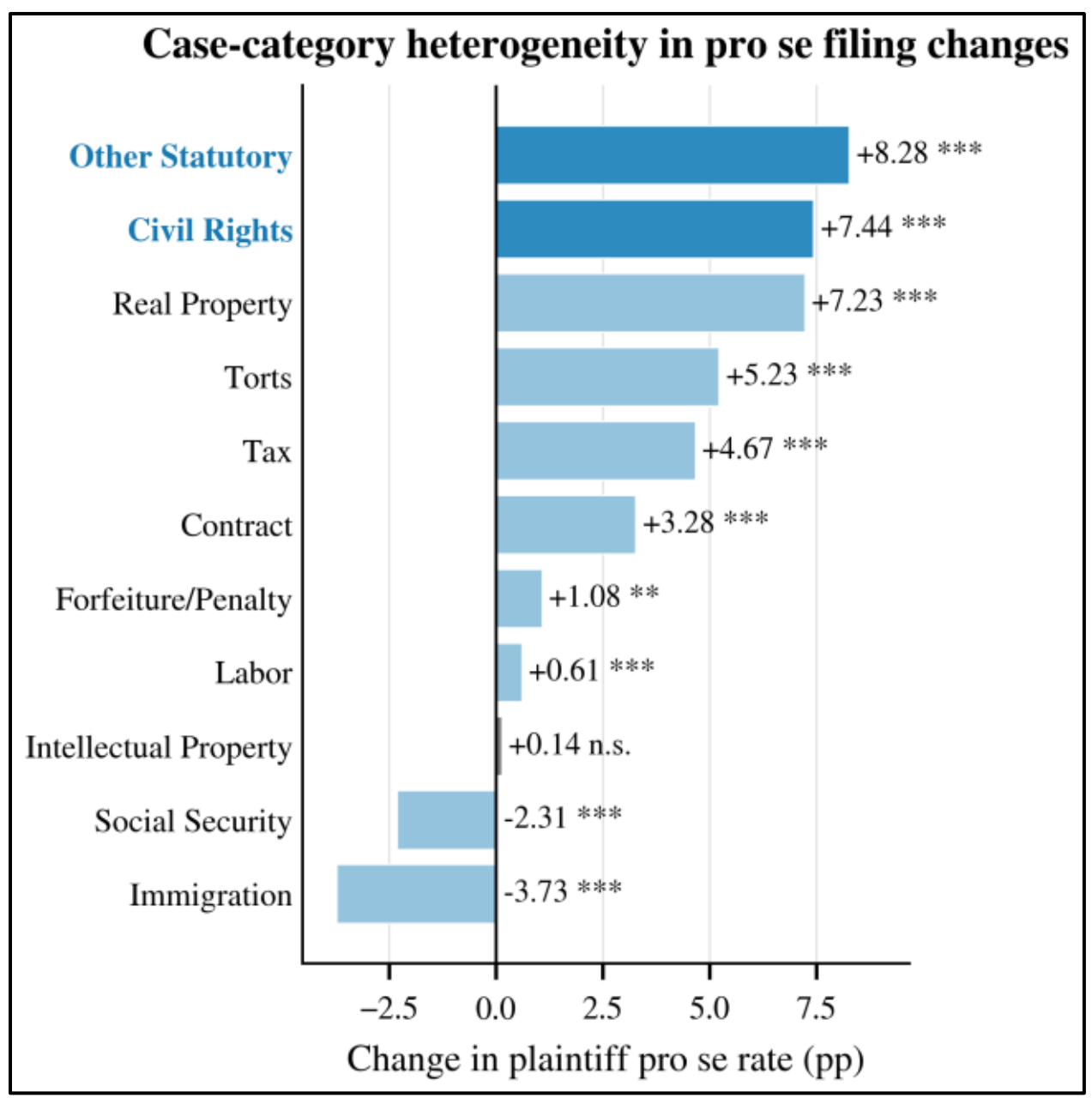

*Figure 3. Pre/post change in plaintiff pro se filing rates across federal civil case categories, FY2008-FY2025, sorted by percentage-point change. Dark blue: focal categories (Civil Rights, Other Statutory); light blue: other statistically significant categories; gray: non-significant. The increase is uneven across the docket, ranging from +8.28 pp (Other Statutory) to declines of -2.31 pp (Social Security) and -3.73 pp (Immigration). Significance: chi-square tests.*

# 5. Constructing AI-Flagged Complaint Subset

This section turns from filing rates to complaint text. The goal is not to prove individual AI use, but to identify complaints whose surface features make AI-assisted drafting especially plausible.

## 5.1 Measuring AI-Consistent Drafting

The AI-consistent drafting measure combines three textual indicators associated with AI-generated prose. (1) Sentence-length variability: AI-consistent text is expected to have lower sentence-length standard deviation (Tang, Chuang, and Hu 2024). (2) AI-marker density: AI-consistent text is expected to use a higher concentration of vocabulary that appears disproportionately in AI-generated writing (Kobak et al. 2025; Liang et al. 2024). (3) Moving-average type-token ratio (MATTR; Covington and McFall 2010): AI-consistent text is expected to show higher lexical diversity via MATTR (Kendro, Maloney, and Jarvis 2026; Fredrick and Craven 2025). Each indicator is converted into a direction-adjusted rank, so that higher values indicate stronger AI-consistent drafting signatures. The combined score averages these three ranked indicators.

A complaint is treated as AI-flagged when its combined score exceeds 2.25, equivalent to an average P75 rank across the three direction-adjusted indicators. The threshold is calibrated against the pre-AI baseline. Thus, the relevant quantity is not the mere existence of high-scoring complaints before public GenAI access, but the net post-AI increase above that structural high-score baseline. The measure is therefore best understood as a proxy for visible AI-consistent drafting, not as a direct detector of actual AI use.

The complaint-text sample shows a substantial post-AI increase in AI-flagged complaints. In the pooled form-and-non-form sample, the net AI-flagging increase is 9.8%. The increase is larger among non-form complaints, where the net post-AI increase is 13.9%. Form complaints show a smaller net increase of 4.3%. Consistent with the rationale described in Section 3, the detectable signal is larger in non-form complaints. The primary RQ2 analyses therefore focus on post-GenAI linked non-form AI-flagged complaints, while form complaints remain a comparison group.

## 5.2 Credibility Checks for the AI Proxy

Because the AI flag is a proxy, we evaluate its credibility through four checks, moving from external timing evidence to internal text-based checks. Each is reported below; the proxy behaves as expected across all four.

First, the AI-flagging pattern aligns with external proxies for public GenAI use and attention. We compare quarterly AI-flagging rates to two independent proxies: reconciled ChatGPT weekly active-user estimates and national Google Trends interest in “AI.” All correlations reported in this paragraph are Spearman correlations. These two external series are themselves strongly aligned in the post-AI window, supporting their use as measures of a common public-adoption and attention trend ($\rho$=+0.991). The AI-flagging series moves with both proxies. In the pooled non-form focal-category sample, the quarterly AI-flagging rate is strongly correlated with ChatGPT weekly active users ($\rho$=+0.936) and Google Trends interest ($\rho$=+0.944). The pattern is similar within each focal category: Civil Rights correlates strongly with both ChatGPT weekly active users ($\rho$=+0.918) and Google Trends ($\rho$=+0.930), as does Other Statutory ($\rho$=+0.927 and $\rho$=+0.937, respectively). This supports the temporal credibility of the proxy: the AI-flagging rate rises when independent proxies show broader public GenAI use and attention increasing.

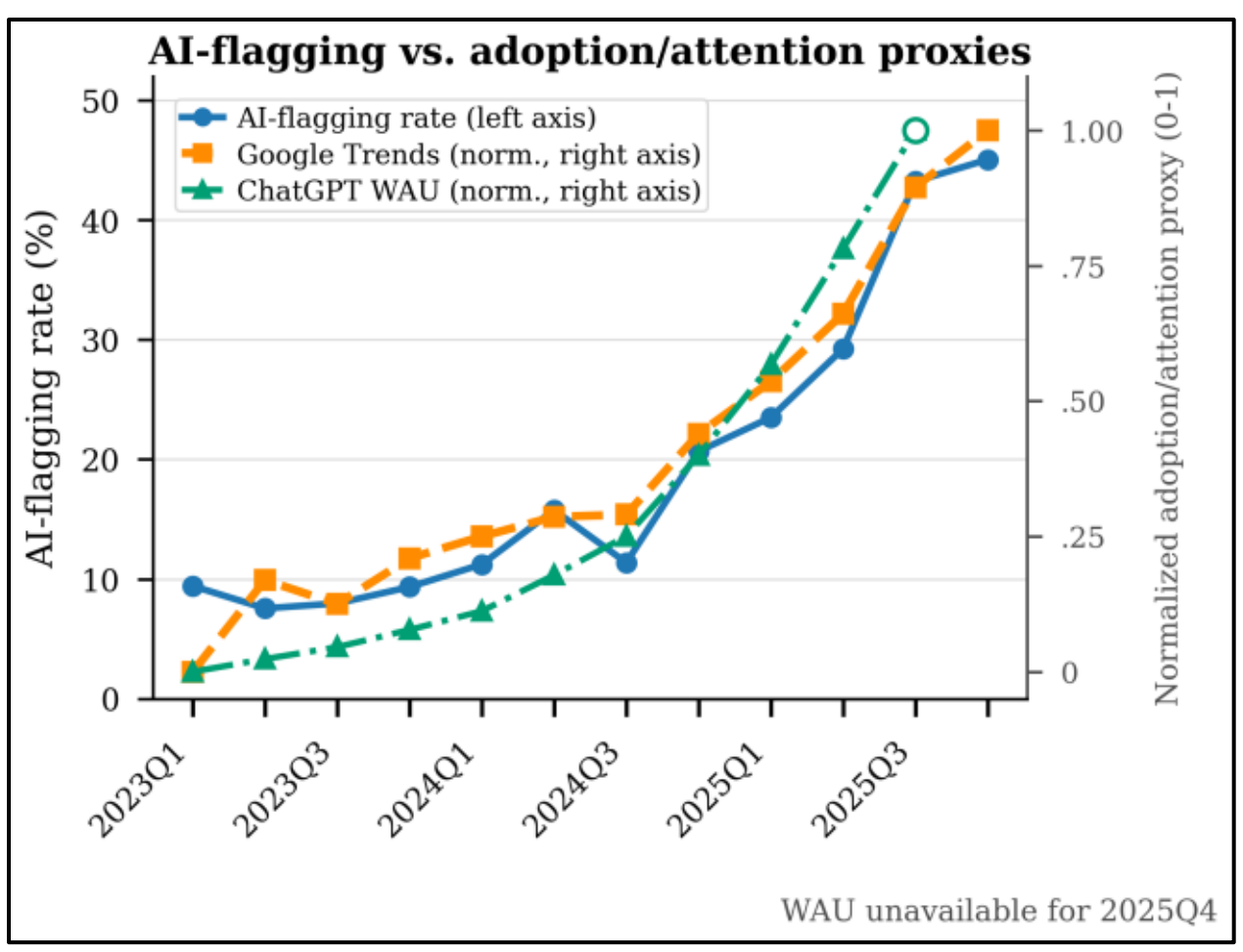

*Figure 4. Quarterly AI-flagging rate for pooled focal-category non-form complaints (2023-Q1 to 2025-Q4) against two min-max normalized GenAI proxies: U.S. Google Trends interest in "AI" and ChatGPT weekly active users (WAU; missing 2025-Q4). Spearman ρ = +0.944 (n=12, Google Trends) and +0.936 (n=11, ChatGPT WAU).*

Second, the component indicators move in the expected directions. In non-form complaints, sentence-length variability falls by 5.8% after the GenAI cutoff, AI-marker density rises by 5.8%, MATTR-50 rises by 1.0%, and the MATTR-100 sensitivity measure rises by 1.6%. All four changes are statistically significant at $p<0.0001$. The full complaint-text sample shows the same directional pattern, though with smaller magnitudes. This supports the internal coherence of the score: the post-AI shift is not driven by a single arbitrary feature.

Third, the indicators increasingly cluster within the same complaints. Using the P75 threshold for each component and requiring all three indicators to exceed their AI-consistent threshold, joint co-occurrence rises from 2.08% pre-AI to 8.31% post-AI. More importantly, the observed-versus-independence multiplier rises from 1.33 to 2.60, with a bootstrapped difference of +1.26 and a 95% confidence interval of [+1.00, +1.52]. This is a strong credibility check because co-occurrence is not used to define the final AI flag. It independently shows that the underlying features increasingly appear together, consistent with a common drafting phenomenon rather than unrelated textual drift.

Fourth, the post-AI signal is not fragile to the exact cutoff. In non-form complaints, the net AI-consistent increase remains positive across alternative thresholds: +15.4% at score >2.00, +13.9% at the primary threshold of >2.25, and +10.3% at score >2.50. As expected, the estimated magnitude decreases as the threshold becomes more conservative, but the post-AI increase remains positive across all thresholds. The proxy passes all four checks, strengthening its credibility as a measure of visible AI-consistent drafting—though not of individual AI use.

## 6. Patterns of AI-Flagged Complaints

This section relies on section 5 to ask how complaints in the AI-flagged subset differ from otherwise comparable complaints. The linkage between the complaint-text sample and the FJC metadata is strong: 12,451 of 12,842 complaint-text records match to FJC records, a 97.0% match rate. The linked complaint sample contains 12,451 cases, including 7,284 pre-AI cases and 5,167 post-AI cases.

### 6.1 Legal Citation Density

Consistent with empirical work treating legal-citation features as meaningful legal-text variables (Tippett et al. 2022), citation density represents the number of legal citations (i.e., case-law, U.S. Code, and Code of Federal Regulations references) per 1,000 words. The citation-density result is strongest in non-form complaints. In that subset, post-AI AI-flagged complaints have a median citation density of 6.01, compared with 3.98 for all pre-AI non-form complaints. This represents a 51.1% increase ($p<0.0001$, $r=+0.234$). The same pattern appears within the post-AI period: AI-flagged non-form complaints have higher median citation density than unflagged non-form complaints, 6.01 compared with 4.40, a 36.6% difference ($p<0.0001$, $r=+0.170$). The pattern is also present in each of the focal case categories: citation density rises by 62.2% in Civil Rights non-form complaints and by 34.0% in Other Statutory non-form complaints.

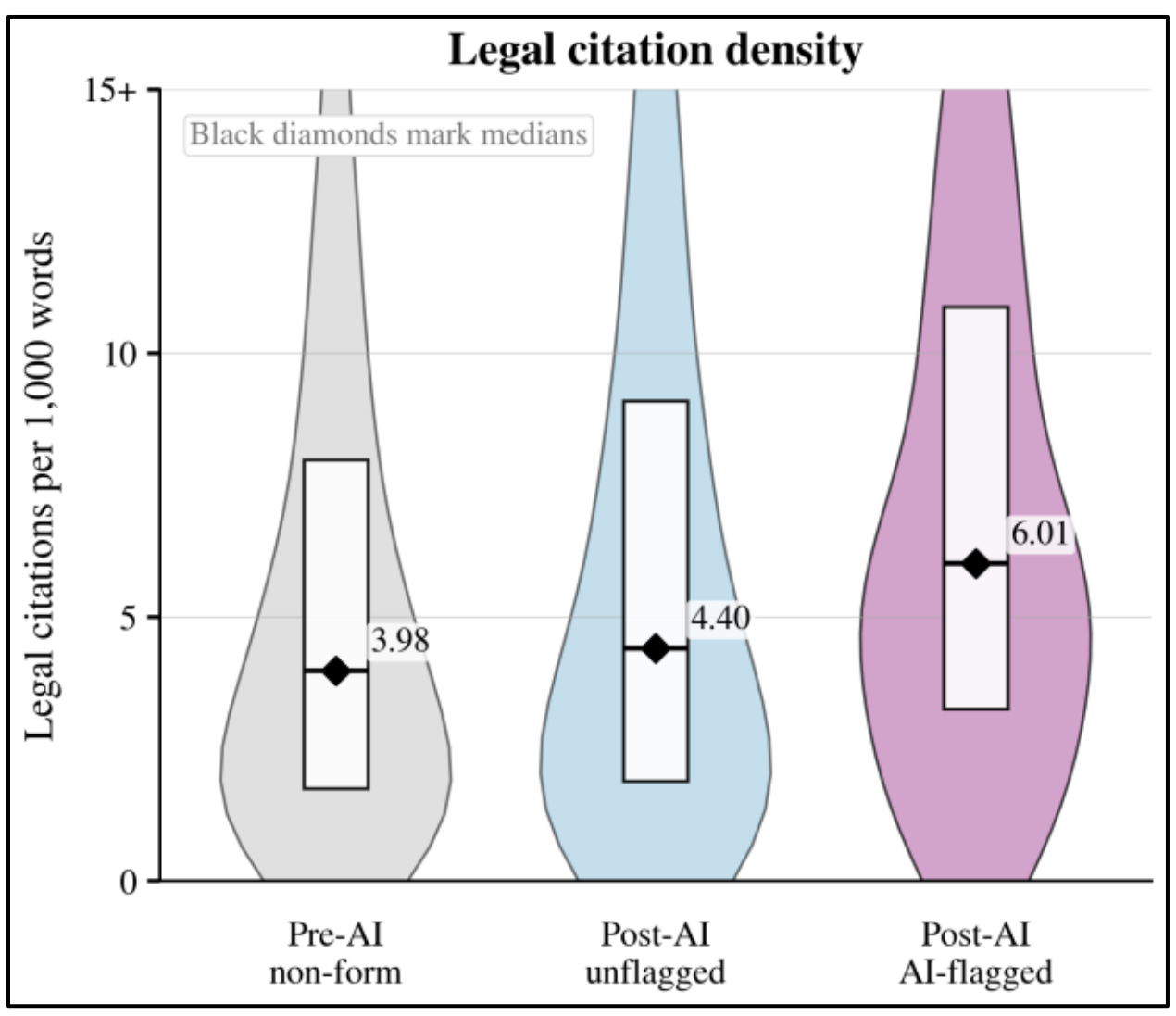

*Figure 5. Legal citation density (citations per 1,000 words) in non-form pro se complaints, FY2018-FY2025, with November 30, 2022 ChatGPT release as the cutoff. Violins show document-level distributions; boxes mark the IQR and diamonds mark medians. Post-AI AI-flagged complaints are markedly denser than both pre-AI complaints (medians 6.01 vs. 3.98; +51.1%, p<0.0001, r=+0.234) and post-AI unflagged complaints (6.01 vs. 4.40; +36.6%, p<0.0001, r=+0.170). Y-axis capped at 15 for readability; statistics use uncapped data.*

Form complaints do not show the same citation-density pattern. In form complaints, post-AI flagged complaints have lower median citation density than both pre-AI all form complaints and post-AI unflagged form complaints: 4.44 compared with 5.08 (−12.5%, p=0.0203, r=−0.088) and 5.79 (−23.3%, p<0.0001, r=−0.188), respectively.

The citation-density result therefore supports a limited empirical claim: AI-flagged non-form complaints are more citation-rich than comparison complaints, while form complaints do not exhibit the same stable pattern. Whether higher citation density corresponds to better litigation performance is addressed below.

## 6.2 Plaintiff Composition

The goal of this subsection is descriptive: to identify whether AI-flagged complaints are concentrated among particular types of litigants or geographies.

**First-time versus repeat filers**. Within the post-AI linked non-form sample, 79.4% of AI-flagged complaints are associated with observed first-time filers, compared with 70.2% of unflagged complaints (+9.3 percentage points; OR=1.638, 95% CI [1.280, 2.097], p<0.001). The result remains essentially unchanged when the mixed launch quarter is excluded in a model-aligned quarterly comparison: 79.4% of AI-flagged non-form complaints and 69.8% of unflagged non-form complaints are observed first-time filings (+9.6 percentage points; OR=1.660, 95% CI [1.294, 2.129], p<0.001). This result matters because it makes the AI-flagged subset less consistent with a pure repeat- or serial-filer explanation. AI-consistent drafting is disproportionately visible among plaintiffs who had not previously appeared in the linked focal sample. The measure is observational: "first-time" means first observed appearance of the primary plaintiff name in the sample, not proof that the person had never filed elsewhere.

**Financial-constraint proxy**. We use in forma pauperis (IFP) status, or fee-waiver status, as a financial-constraint proxy (Hammond 2019). IFP status allows a plaintiff who files a declaration of financial inability under penalty of perjury to file their complaint without prepaying the filing fee (28 U.S.C. § 1915(a)(1)).

The fee-waiver evidence does not show a meaningful association with AI flagging. In the post-AI non-form subset, fee-waiver status appears in 27.3% of AI-flagged complaints and 24.4% of unflagged complaints, but it is not statistically significant (p=0.12). At the population level, the post-AI pro se filing surge occurs while pro se fee-waiver rates fall, from 42.3% pre-AI to 35.9% post-AI (−6.35 percentage points, p<0.001). Thus, the evidence does not support treating AI-flagged complaints as primarily a fee-waiver or indigent-litigant phenomenon. The cautious conclusion is that fee-waiver status is not meaningfully or consistently associated with AI flagging.

**Gender**. Name-inferred gender provides a more tentative composition signal. We use NamSor's name classifier and treat gender as unresolved when the calibrated probability falls below the threshold. At the primary threshold (>= 0.80), AI-flagged post-AI non-form complaints have a modestly higher inferred-female share among binary-classified names than unflagged non-form complaints: 19.7% versus 15.0% (+4.6 percentage points; OR=1.389, 95% CI [1.002, 1.927], p=0.051). The direction and magnitude are similar under the secondary 0.70 threshold (26.6% versus 22.0%, +4.6 percentage points; OR=1.286, 95% CI [0.993, 1.666], p=0.058). Because gender is name-inferred rather than self-reported, and because the pooled non-form comparison is marginal under both thresholds, we treat this result as suggestive rather than conclusive evidence of gender differences.

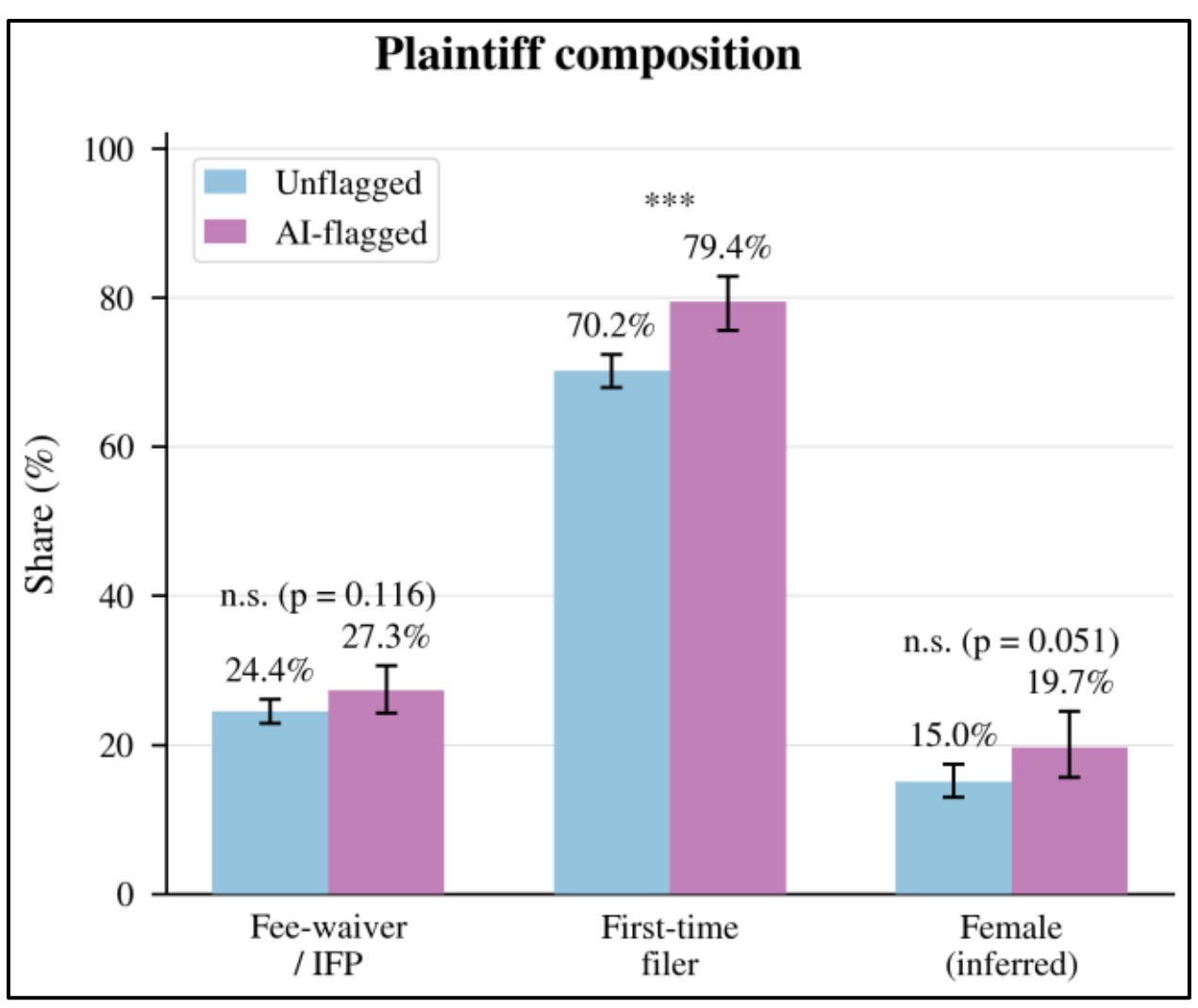

*Figure 6. Plaintiff composition for AI-flagged versus unflagged non-form pro se complaints in the post-GenAI focal sample (Civil Rights and Other Statutory). Bars: shares with 95% Wilson binomial CIs. First-time filer = primary plaintiff's first observed appearance in the FY2018-FY2025 identity-linked sample, not an absolute first-ever federal filing. Name-inferred female uses NamSor (calibrated probability ≥ 0.80); denominator is binary-classified primary-plaintiff names only. AI-flagged complaints are disproportionately first-time filers (+9.3 pp); IFP and female-share differences are non-significant (two-proportion z-tests).*

**Geographic distribution**. AI-flagged complaints are geographically uneven across federal circuits. In the post-AI linked non-form sample, the highest AI-flag rates appear in the Tenth Circuit (33.3%), Fourth Circuit (30.0%), First Circuit (29.3%), D.C. Circuit (29.1%), and Eleventh Circuit (24.7%). Lower rates appear in the Ninth Circuit (14.9%) and Sixth Circuit (15.5%). A concentration test confirms that the non-form AI-flagged complaints are not evenly distributed across circuits ($\chi^2$=55.319, df=11, p<0.001, Cramér's V=0.132).

## 6.3 Procedural Trajectories and Case Outcomes

The final set of analyses compares procedural progress and outcomes. These analyses are limited to terminated cases and therefore the primary post-AI outcome working set is smaller (n=3,924). That design allows comparison between AI-flagged and unflagged complaints within the same post-AI observation window, but it does not eliminate right-censoring concerns: many recent post-AI cases remain pending, and therefore dismissal, procedural-progress, and win-rate results should be interpreted as terminated-case analyses rather than final outcomes for all post-AI cases.

**Procedural progress**. Procedural progress is measured using the FJC's procedural-progress classifications, which capture where a case ends rather than how long it took. 'Code 02' marks cases terminating before issue is joined (an early stage, before the defendant answers); 'Code 04' marks cases terminating after issue is joined through judgment on motion (a more advanced stage).

In the post-AI linked non-form subset, AI-flagged complaints are more likely to terminate at Code 02 than the unflagged comparison group. Code 02 appears in 47.78% of AI-flagged non-form complaints, compared with 39.04% of unflagged non-form complaints, a +8.75 percentage-point difference (FDR-adjusted p=0.002). Code 04 moves in the opposite direction: 10.84% of AI-flagged non-form complaints and 13.19% of unflagged non-form complaints terminate at Code 04, a -2.35 percentage-point difference, however, it is not statistically significant (p=0.198). Among dismissal-coded terminations only, Code 02 remains higher for AI-flagged non-form complaints, 60.9% compared with 53.9% for unflagged non-form complaints (+7.0 percentage points, p=0.047).

The procedural-progress evidence therefore points to a specific pattern: AI-flagged non-form complaints are more likely to terminate at an earlier procedural stage.

**Dismissals**. We define dismissals as three FJC dispositions—Want of Prosecution, Lack of Jurisdiction, and Other Dismissal (a residual category). The definition excludes voluntary dismissals and settlements.

In the primary non-form comparison, AI-flagged complaints have a higher dismissal rate than the unflagged comparison group. Among post-AI terminated non-form cases, 61.1% of AI-flagged complaints are dismissed, compared with 53.6% of unflagged complaints, a +7.5 percentage-point difference (p=0.006). The component decomposition clarifies what drives the non-form dismissal gap. 45.3% of AI-flagged non-form complaints terminate through Other Dismissal, compared with 40.0% of unflagged non-form complaints (p=0.049). Thus, the dismissal gap in the primary comparison is concentrated mainly in the residual Other Dismissal category, with a smaller and less precise jurisdictional difference.

**Pro se win rate**. Pro se win rate is measured as plaintiff judgments divided by plaintiff-plus-defendant judgments among cases with clear judgment coding. The results do not show a plaintiff-win advantage for AI-flagged complaints, and the observed differences are not statistically significant. In the full post-AI linked set, the win rate is 3.7% for AI-flagged complaints and 5.3% for unflagged complaints, (p=0.525). In the primary non-form comparison, the win rate is 6.4% for AI-flagged complaints and 8.4% for unflagged complaints (p=0.635).

Plaintiff wins are relatively rare in the pro se data. The appropriate conclusion is therefore limited: the evidence neither shows that AI-flagged complaints have improved pro se win rates nor supports a claim that AI flagging worsens them.

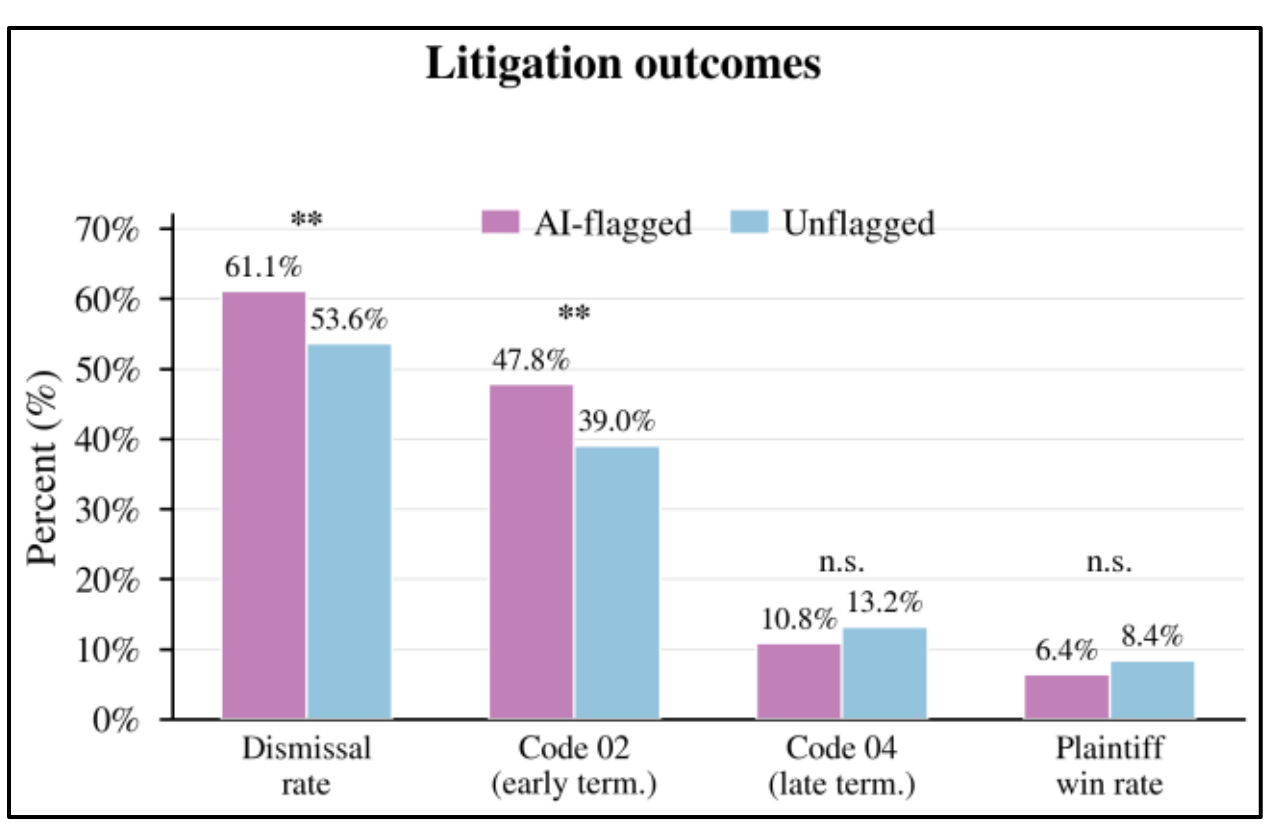

*Figure 7. Litigation outcomes for AI-flagged vs. unflagged post-GenAI linked non-form complaints in the focal categories. Pink: AI-flagged; light blue: unflagged. AI-flagged complaints are more often dismissed (61.1% vs. 53.6%, +7.5 pp) and more often terminate early (Code 02; 47.8% vs. 39.0%, +8.8 pp), with no plaintiff win-rate advantage. Code 02: termination before issue is joined; Code 04: termination after issue is joined through judgment on motion. Dismissal and Code 02/04 rates use terminated cases (n=406; n=1,911); plaintiff win rate uses cases with clear plaintiff/defendant judgment coding (n=47; n=333). Pairwise chi-square tests.*

# 7. Discussion

## 7.1 Access to Justice or Access to Drafting?

One interpretation of the post-AI filing surge is access-enhancing. If GenAI lowers the time, cost, and expertise required to produce a complaint, then some litigants who could not practically file in a non-GenAI world may now be able to initiate litigation without counsel. This possibility tracks a prominent promise in the legal-AI and access-to-justice literature: that generative AI and related legal technologies may reduce the resources required to obtain basic legal help or produce legal documents (Doyle 2025).

This framing, however, requires a distinction the access-to-justice literature has long pressed: between formal entry into the court system and the substantive ability to obtain meaningful legal relief. GenAI may expand access to drafting — the ability to generate a complaint that can be filed in court — without expanding effective access — the ability to survive threshold review and obtain a meaningful legal outcome (Sandefur 2019). Whether the post-AI surge actually translates expanded drafting access into improved litigation outcomes is the question Section 7.2 takes up, and it complicates the simple access-enhancing story.

However, even setting that question aside, access to drafting carries real value of its own. For some litigants, the ability to file may itself be meaningful. Participation can have democratic, dignitary, and procedural-justice value, distinct from whether the plaintiff ultimately prevails (Lind and Tyler 1988; MacCoun 2005; Tyler 2006). Being heard, having a complaint docketed, and engaging with the legal system as a participant rather than as someone whose grievance never enters it are outcomes the access-to-justice literature has long treated as constitutive of access, not as consolation for losing on the merits. A litigant whose claim is dismissed is in a different position than a litigant who could never reach a court at all.

There is also a more straightforwardly remedial point. Win rates and filing volumes are distinct quantities. If GenAI enables more pro se plaintiffs to file while plaintiff-favorable judgment rates remain stable, then more plaintiffs may obtain plaintiff-favorable judgments in absolute terms. Expanded drafting access can produce additional successful outcomes through volume alone, without any per-case efficacy gain.

## 7.2 The Litigation-Efficacy Paradox

Section 7.1 left open whether expanded drafting access has actually translated into improved litigation outcomes. The data suggest it has not. One might expect AI-flagged complaints to perform better than unflagged complaints, because if these complaints reflect GenAI assistance, they would suggest that litigants had access to a drafting resource they previously lacked. Consistent with that expectation, AI-flagged complaints do appear more formally supported than unflagged complaints in terms of citation density. Yet on litigation outcomes, the pattern reverses: AI-flagged non-form complaints are associated with higher dismissal rates, earlier procedural termination, and no observed win-rate advantage. More polished-looking complaints perform worse in court. What might explain this litigation-efficacy paradox?

One possible explanation is surface-level assistance. GenAI may help litigants produce complaints that contain more citations and appear more legally formal without improving the legal merit of the underlying case. Federal pleading doctrine favors factual plausibility over legal formality (Bell Atlantic Corp. v. Twombly 2007; Ashcroft v. Iqbal 2009), and empirical work has documented that this plausibility regime has produced systematically higher dismissal rates for pro se and civil-rights plaintiffs (Hatamyar Moore 2010; Gelbach 2012; Reinert 2015). LLMs may be well-suited to the former and less so to the latter — generating fluent, citation-dense legal prose is something LLMs can do, while constructing a fact-specific narrative grounded in the litigant's own circumstances that supports a plausible legal claim is a task on which they perform less reliably (Katz et al. 2024; Dahl et al. 2024, Cohen-Sasson 2026).

A complementary explanation concerns not the content of the complaint, but the selection of cases that get filed. If GenAI reduces the practical burden of drafting, it may enable people to file cases they otherwise would not have filed. Some of those cases may be meritorious; others may be marginal, premature, jurisdictionally defective, or legally weak. As the cost of producing a complaint falls, legal merit may play a smaller role in the threshold decision to file. Under this account, the filing surge reflects expanded capacity to file, not necessarily improved screening of which claims should be filed (Priest and Klein 1984; Bebchuk 1996).

This case-selection account need not assume that litigants knowingly file weak claims. A related mechanism is overconfidence potentially reinforced by sycophancy. Sycophancy refers to the tendency of LLMs to agree with, validate, or amplify a user's stated beliefs or preferences, even when doing so may produce inaccurate outputs (Sharma et al. 2024). If an LLM validates the user's framing, a non-expert litigant may come to see a weak or marginal claim as more viable than it is (Buçinca, Malaya, and Gajos 2021; Sun et al. 2025). Sycophancy could help explain why lower drafting barriers might translate into more weak filings: not (only) because filing becomes easier, but because GenAI may also make filing feel legally justified for a pro se litigant.

### 7.3 Uneven Diffusion and Distributional Limits

The results also suggest that GenAI-associated change in pro se litigation is unevenly distributed. The AI-consistent drafting pattern varies across complaint type and geography, and the composition evidence points to a more nuanced access story than a simple account of broad democratization. The observed first-time result is the strongest composition signal. AI-flagged complaints are not simply concentrated among repeat litigants who already knew how to navigate federal court. Instead, AI-consistent drafting appears disproportionately among plaintiffs newly observed in the linked focal sample. This finding supports a more specific version of the access-to-drafting claim: the post-AI shift may include new entry into the focal pro se docket, not only increased activity by repeat filers.

Gender points in a similar direction, but more weakly. The name-inferred results show a modestly higher inferred-female share among AI-flagged complaints, but the result is marginal. We therefore treat the gender pattern as suggestive rather than conclusive evidence of gender differences.

The fee-waiver evidence places an important limit on this access-to-drafting interpretation. Access-to-justice scholarship often focuses on financially constrained litigants (Legal Services Corporation 2022). Within the post-AI period, AI-flagged complaints have a 2.9 pp higher IFP share than unflagged complaints (27.3% vs 24.4%), but the difference is not statistically significant; and at the population level, pro se fee-waiver rates fall from 42.3% to 35.9%. Neither pattern supports treating AI-assisted drafting as primarily a fee-waiver phenomenon. If GenAI is helping some litigants file, the observed increase does not appear to be driven by a rising share of fee-waiver litigants. The benefits of AI-assisted filing capacity may therefore be distributed among pro se litigants more broadly, rather than concentrated among those facing the most acute financial barriers to court access. This aligns with broader technology diffusion patterns: availability is not the same as equitable uptake or effective use (DiMaggio and Hargittai 2001; Hargittai 2002; Humlum and Vestergaard 2025).

The overall picture is therefore one of uneven diffusion. The post-AI shift reaches new entrants to the federal civil pro se docket — most clearly first-time filers, and tentatively, a higher share of name-inferred female plaintiffs — but it is not concentrated among the financially constrained populations that scholars have historically focused on.

### 7.4 Court Burden and Asymmetric Adaptation

The findings raise institutional-capacity questions for courts (Stienstra, Bataillon, and Cantone 2011). Because AI-flagged complaints are more often dismissed and terminate at earlier stages, courts may be especially exposed to threshold-stage review work.

The broader issue is asymmetric adaptation. To begin with, it is not necessarily desired—from a normative standpoint—that courts should or want to use AI. But even if we assume they are interested in adopting AI, there are structural differences between courts and pro se litigant. Litigants can adopt public GenAI tools quickly because they are widely available, relatively low-cost, and easy to access. Courts, by contrast, may adopt AI more slowly because of procurement requirements, ethics standards, reliability concerns, and so forth (New York State Bar Association 2024). If litigants' drafting capacity expands faster than courts' institutional capacity to process and screen the resulting filings, legal institutions may experience a capacity mismatch, further exacerbating any friction related to the prospects of GenAI in pro se litigation (Conference of State Court Administrators 2024).

The implication is not that pro se litigants should be prohibited from using AI. Such a rule may be difficult to monitor, unevenly enforceable, and normatively unattractive if AI helps some people secure legal remedies they could not otherwise pursue. The better implication is that court-side adaptation is required (Engstrom and Gelbach 2021). Possible responses include staff training, triage protocols, policies for AI-assisted filings, and careful evaluation of court-side AI tools (Conference of State Court Administrators 2024; Re and Solow-Niederman 2019).

## 8. Conclusion

The paper asked whether public access to generative AI tools is associated with measurable changes in federal civil self-representation. The macro analysis shows that it is: the federal civil plaintiff pro se rate rose from 11.33% to 16.94% (+5.61 percentage points), a shift that interrupted time-series evidence describes as an accelerating post-AI trend rather than a one-time level change, and that is concentrated in specific fields, particularly Civil Rights and Other Statutory cases. The complaint-level analysis, built on a pre-AI-calibrated text proxy that passes four credibility checks, shows that a net 13.9% of post-AI non-form complaints carry AI-consistent drafting signatures. These flagged complaints are markedly more citation-dense than the unflagged comparison group, yet are more likely to be dismissed, more likely to terminate at earlier procedural stages, and show no win-rate advantage. They are also compositionally distinctive: AI-flagged complaints are disproportionately associated with observed first-time filers, and show a modest, suggestive increase in name-inferred female plaintiffs. Together, the empirics support a distinction the paper develops throughout: GenAI appears to expand access to drafting without a matching expansion of access to legal remedies, producing a litigation-efficacy paradox in which more formally polished complaints fare worse in court.

These findings provide large-scale empirical evidence for a claim signaled at the outset: public-facing GenAI tools can reshape legal institutions from the bottom up, through litigants, without any formal court adoption of AI. This is a litigant-side mechanism of institutional change, distinct from the lawyer- and court-side debates that have dominated AI-and-law scholarship. For access-to-justice research, the findings complicate the optimistic strand of the literature: drafting access expanded, but the composition of who appears to benefit — disproportionately observed first-time filers, a modest and suggestive increase in name-inferred female plaintiffs, and no concentration among fee-waiver litigants — reframes both who is helped and what they receive. The result reinforces the long-standing distinction between formal entry to court and substantive access to remedy. For court administration, the asymmetric adaptation problem becomes empirically grounded: filings and threshold-screening burdens are rising faster than courts can absorb them, and the implied response is court-side adaptation rather than litigant restriction. For AI-and-society scholarship more broadly, this is one of the first concrete institutional settings where the diffusion effects of public LLM access can be measured at scale; the pattern documented here — broader formal participation, unevenly distributed, with no corresponding gain in substantive outcomes — may be a useful template for other domains where GenAI lowers production cost without improving the quality of the underlying claim (Dell'Acqua et al. 2026; Liu et al. 2024).

Several questions remain open. First, the outcome analyses depend on terminated cases and are subject to right-censoring, as many post-AI cases remain pending; longer observation windows will permit re-estimation of dismissal, procedural-progress, and win-rate gaps, and will help distinguish persistent effects from artifacts of early-terminating cases being over-represented in the current window. Second, the AI-consistent drafting proxy captures direct, visible drafting signatures and is conservative about indirect AI uses such as brainstorming, translation, legal research, or light editing; litigant-side validation through surveys, interviews, or court-clerk reports would help bound the gap between visible signatures and total AI use, and would test whether some unflagged complaints are AI-assisted in less detectable ways. Third, the candidate mechanisms proposed in Section 7 — surface-level formality under plausibility pleading, expanded marginal-case selection, and sycophancy-reinforced overconfidence — cannot be separated with this study's design; each calls for a different method, from doctrinal analysis of dismissal opinions, to merit-coding of a complaint subset, to experimental work with LLM systems and lay users. Finally, future work should extend the inquiry beyond federal pro se plaintiffs to pro se defendants, where threshold dynamics differ; to state civil courts, which carry the majority of pro se litigation and where access-to-justice stakes are arguably higher (Shanahan et al. 2022); and to the institutional adaptations courts develop — triage protocols, staff training, AI-assisted filing rules — as litigant filing capacity continues to outpace court capacity.